\documentclass[doublecol]{epl2} 

\usepackage{amsmath}
\usepackage{etoolbox}
\AtBeginEnvironment{eqnarray}{\setlength{\arraycolsep}{2pt}}

\newcommand{\be}{\begin{equation}}
\newcommand{\ee}{\end{equation}}
\newcommand{\wdg}{\wedge}   
\newcommand{\ot}{\otimes}
\newcommand{\fant}[1]{\phantom{#1}}

\title{Abbott-Deser energy in connection with \\Thirring's superpotentials}
\shorttitle{Abbott-Deser Energy} 

\author{Ahmet Baykal\inst{1} \and Tekin Dereli\inst{2} }
\shortauthor{A. Baykal,  T. Dereli}

\institute{                    
  \inst{1} Ni\u gde \"Omer Halisdemir University, Faculty of Arts and Sciences, Department of Physics, Bor yolu \"uzeri, Merkez kamp\"us, 51240 Ni\u gde, TURKEY\\
  \inst{2} Department of Physics, College of Sciences, Ko\c{c} University,  34450 Sar\i yer, \.Istanbul, TURKEY
}

\abstract{
We linearize vacuum Einstein field equations with a cosmological constant around a curved background to elaborate on the reconstruction of the  Abbott-Deser charges 
and incorporate a spin connection into the definition using the algebra of differential forms on a given curved background spacetime.
}

\begin{document}

\maketitle

\section{Introduction}

There are  many definitions of total energy in general theory of relativity, each adapted to some particular cases of interest.  It is desirable to 
disclose their interrelationship and differences.  Previously, Wallner and Thirring \cite{thirring1,wallner-thirring2} explicitly showed that using the expression for  the Thirring 2-forms, it is possible to obtain sundry energy definitions of energy, given by for example, Freud \cite{freud},  M\o ller \cite{moller1,moller2}, and Landau and Lifschitz \cite{ll}. See also the recent paper by B\"ohmer and Hehl \cite{Bohmer-Hehl}.
Here, we elaborate on the reformulation of the Abbott-Deser energy definition  in terms of differential forms  that helps to gain further insight into complicated mathematical expressions. The construction of the Abbott-Deser (AD) charges for generic gravitational models depends on the background Killing vector fields 
\cite{abbott-deser,tekin-rev,cebeci-sarioglu-tekin1,deser-tekin1,deser-tekin2}.

There are two essential ingredients for the construction of Abbott-Deser charges:
\begin{itemize}
\item[(i)]
Background Killing vector fields defined for any gravitational model having general coordinate covariance and local Lorentz invariance, 
\item[(ii)]
Linearization of the field equations around a background and obtain a flux integral obtained by making use of Stokes' theorem.
\end{itemize}

  In particular, Killing energy is constructed  with the help of an asymptotically timelike Killing vector field. To facilitate comparison,  we adopt the  same conventions used in \cite{cebeci-sarioglu-tekin1} and refer to Appendix A therein for the derivation of the linearized quantities using an orthonormal coframe for the full field equations and the curved background as well. The extraction of tensor components linearized around a curved background using the spin connection formalism can be found in \cite{cebeci-sarioglu-tekin2}, extending the discussion of the results in \cite{cebeci-sarioglu-tekin1}.

We first briefly discuss the linearized quantities for convenience.
For a sufficiently well behaved perturbation 1-form $\phi_a\equiv \phi_{ab}\bar{e}^b$, the metric tensor can be written in the form
\be
g=\eta_{ab}e^a\ot e^b=\bar{g}+\phi_{a}\ot \bar{e}^a+\bar{e}^a\ot\phi_{a}+O(\phi^2)
\ee
to the first order in $\phi_a$'s.
The barred quantities refer to an orthonormal coframe of  a curved  background 
\be\label{coframe-expansion-def}
e^a
=
\bar{e}^a+\phi^a+O(\phi^2)
\ee
where 
$
\phi^a
=
\phi^a_{\fant{a}b}\bar{e}^b
$
with $\phi_{[ab]}=0$. $\phi_a$ can be considered as  covector valued-1-forms in the background with the metric $\bar{g}$ if one is confined to first order quantities  for a given background \footnote{Throughout the paper a beginning Latin index refers to an 
orthonormal  (co)frame basis, $a, b, c\ldots=0, 1, 2, 3$, whereas a Greek index  refers to a  coordinate (co)frame basis $\mu, \nu,\alpha, \ldots=0, 1, 2, 3$ and  spatial indices are labeled using the middle Latin letters $i, j, k\ldots$.}. 
The expansion of the basis coframe  1-forms in the form given in Eq. (\ref{coframe-expansion-def})
is akin to the familiar expansion of the form $x=\bar{x}+\varepsilon x_{(1)}+O(\varepsilon^2)$ in terms of a suitable expansion parameter $\varepsilon$. In our case it is assumed that $\phi_{ab}$s are  well behaved functions that allow one  to introduce a perturbation expansion around a curved background.
With the help of the expansion in Eq. (\ref{coframe-expansion-def}), and assuming that the Cartan's structure equations are also satisfied for the 
(barred) background coframe and connection 1-forms, Cartan's first structure equations reduce to
\be\label{se-01}
d\phi^a+\bar{\omega}^{a}_{\fant{a}b}\wdg \phi^b+\bar{\omega}^{a}_{(1)b}\wdg \bar{e}^b=0
\ee
to first order in the perturbation 1-forms. Here, $\bar{\omega}^{a}_{(1)b}$ denote the connection 1-forms to the first order part of the linear 
connection $\bar{\omega}^{a}_{L\;b}$, that can be given by
\be
\bar{\omega}^{a}_{(1)b}
=
\bar{\omega}^{a}_{L\,b}-\bar{\omega}^{a}_{\fant{a}b}.
\ee
In terms of background covariant exterior derivative $\bar{D}$, as a set of first order equations, this can be rewritten in a more convenient form
\be\label{reduced-se1}
\bar{D}\phi^a+\bar{\omega}^{a}_{(1)b}\wdg \bar{e}^b=0
\ee
which may be regarded as a system of 2-form equations in the background. Subsequently, (\ref{se-01}) can be inverted to express $\bar{\omega}^{a}_{(1)b}$ 
in terms of the covariant exterior derivative and one ends up with
\be\label{reduced-se11}
\bar{\omega}^{a}_{(1)b}
=
\frac{1}{2}\bar{i}^a\bar{i}_b\left(\bar{D}\phi^c\wdg \bar{e}_c\right)
-
\bar{i}^a\bar{D}\phi_b
+
\bar{i}_b\bar{D}\phi^a
\ee
where $\bar{i}^a$'s are the background contraction operators. The  linearized form of the contraction operators  are $i^L_a=\bar{i}_a-\phi^{b}_{\fant{a}a}\bar{i}_b$.
With the previous assumption $\phi_{[ab]}=0$, one can show that  the first term on the right-hand side vanishes, thus (\ref{reduced-se1})  leads to
\be
\omega^{a}_{L\;b}
=
\bar{\omega}^{a}_{\fant{a}b}
+
\Delta^{a}_{\fant{a}b}
\ee
where the antisymmetric tensor-valued 1-forms $\Delta_{ab}=-\Delta_{ba}$ belong to the background geometry with the explicit expression
\be
\Delta_{ab}
\equiv
-
\bar{i}_a\bar{D}\phi_b
+
\bar{i}_b\bar{D}\phi_a.
\ee
Consequently, using  Cartan's second structure equations, the linearized curvature 2-forms take the form
\be\label{lin-curv-form}
\Omega^{ab}_{L}
=
\bar{\Omega}^{ab}+\bar{D}\Delta^{ab}.
\ee
Note that, the linearized first Bianchi identity $(\Omega^{a}_{\fant{b}b}\wdg e^b)^L=0$ ensures that the index symmetries/antisymmetries 
are retained  to the linear order in $\phi^a$ and this requires $\phi_{[ab]}=0$.
In terms of components, $\Omega^{ab}_{L}$ refer to a composite quantity that is to be expanded into a background zeroth order quantity 
and a term linear in $\phi$ by definition. Keeping this in mind, it is explicitly given by
\be
\Omega^{ab}_{L}
=
\bar{\Omega}^{ab}
+
\frac{1}{2}\left(
R^{ab}_{(1)cd}
+
\phi_{mc}\bar{R}^{m\fant{c}ab}_{\fant{m}d}
-
\phi_{md}\bar{R}^{m\fant{c}ab}_{\fant{m}c}
\right)\bar{e}^{cd}.
\ee
Using this expression in conjunction with (\ref{lin-curv-form}), it is possible to obtain the first order  Riemann tensor
$R^{ab}_{(1)cd}$ in terms of the derivatives of the perturbation 1-forms $\phi^a$ that is identical to the one obtained 
using  an approximate coordinate coframe. This explicit expression for it can be written concisely as an equation for  2-forms,
\be
\frac{1}{2}
R^{(1)}_{abcd}
\bar{e}^{cd}
=
-\phi_c\wdg \bar{i}^c\bar{\Omega}_{ab}+\bar{D}\Delta_{ab}.
\ee
 The brief discussion of the linearization formula above allows one to calculate any linearized quantity.
In particular, the linearized Einstein forms 
\be
(*G^a)^L
=
-
\frac{1}{2}
\left(
\Omega_{bc}\wdg *e^{abc}
\right)^L
=
-
\frac{1}{2}
\Omega^L_{bc}\wdg (*e^{abc})^L
\ee
can be put in the following remarkable form
\be
(*G^a)^L
=
\bar{*}\bar{G}^a_{}+\bar{D}\bar{*}\bar{\mathcal{F}}^a
\ee
where 
\be\label{gen-thirring-form1}
\bar{*}\bar{\mathcal{F}}^a
=
-\frac{1}{2}\Delta_{bc}\wdg \bar{*}\bar{e}^{abc}.
\ee
The vector-valued 2-forms $\bar{*}\bar{\mathcal{F}}^a$ above can be regarded as a generalization of Thirring 2-forms 
to a curved background and can be rewritten in even a more suggestive form 
\be\label{gen-thirring-form2}
\bar{*}\bar{\mathcal{F}}^a
=
\bar{e}^{b}\wdg \bar{*}\left(\bar{e}^a\wdg \bar{D}\phi_b\right).
\ee
The linearized expressions for the Einstein forms in Eqs. (\ref{gen-thirring-form1}) or (\ref{gen-thirring-form2}) are the main results 
of our discussion. As the ADM energy definition can conveniently be expressed  in terms of  Thirring 2-forms in an asymptotically flat background, 
the expressions in (\ref{gen-thirring-form2}) provide the missing link  between the Abbot-Deser definition of total energy and the Thirring 2-form
for the curved background spacetimes.

 We now turn to a brief discussion of the AD construction involving background Killing vector fields.
Let us assume that the gravitational Lagrangian density $n$-form $L=L[g,\lambda]$ depends on the metric tensor whose field equations can be derived from coframe variations of the form
\be
*E^a
\equiv \frac{\delta L}{\delta e_a}
\ee
where $E_a\equiv E_{ab}e^b$ are the covector-valued 1-forms and $\lambda$ denotes the cosmological constant.  

The general coordinate covariance of the field equations requires that the metric field equations are covariantly constant
$D*E^a=0$. In addition, the local Lorentz covariance leads to the identity
\be
e^a\wdg *E^b-e^b\wdg *E^a=0
\ee
that can also be expressed as symmetry conditions $E_{[ab]}=0$. The symmetry property of $\phi_{ab}$ is required for the first Bianchi identity to be satisfied to first order in $\phi_a$. These important identities are valid under fairly general assumptions, that are also  desirable from the physical point of view, for any gravitational Lagrangian. Then, one also has 
\be
D(\xi_a*E^a)
=
\frac{1}{2}E_{ab}\left(D\xi^a\wdg*e^b+D\xi^b\wdg*e^a\right)
\ee
where the right-hand side vanishes identically for a Killing vector field satisfying the Killing identity 
\be\label{killing-eqn}
i_X\nabla_Y \tilde{\xi}+i_Y\nabla_X \tilde{\xi}=0
\ee
for arbitrary vector fields $X, Y$ and $\tilde{\xi}=\xi_ae^a$ is the Killing 1-form associated with the Killing vector field $\xi=\xi^\mu\partial_\mu$.
Consequently, the expression (\ref{killing-eqn}) can be used to define the conserved currents
\be\label{integrand}
Q^a(\xi)
=
\frac{1}{4\Omega_{n-2}G_n}\int_{U} (\xi^aE_{ab}*e^b)^{L}
\ee
with the integral evaluated  at spatial infinity.
The normalization factor multiplying the integral incorporates the Newtonian constant in $n$ spacetime dimensions, namely  $G_n$, and the  
area of the unit sphere $S^{n-2}$ denoted by $\Omega_{n-2}$.
 All the terms in the integrand in (\ref{integrand}), in particular the $(n-1)$-forms $\bar{*}\bar{e}^b$, are  to be restricted to a $(n-1)$-dimensional spatial submanifold $U\subset M$.

The second step is to convert the integrand in (\ref{integrand}) into  exact $(n-1)$-form to define an $(n-2)$-dimensional flux integral at 
spatial infinity $\partial U$, simply by making use of Stokes' theorem for differential forms. Following the original construction, the only technically demanding 
part of the construction that remains is to construct  a 2-form with components $\mathcal{F}_{\mu\nu}$ from the defining relation 
$\xi^\mu E_{\mu\nu}=\nabla_{\mu}\mathcal{F}^{\mu}_{\fant{a}\nu}$,  up to a suitably normalized constant. Having achieved this step, the general expression (\ref{integrand}) for the conserved charges reduces to \be\label{general-formula1}
Q^a(\xi)
=
\frac{1}{4\Omega_{n-2}G_n}\int_{\partial U} dS_i\mathcal{F}^{ia}.
\ee
Note that in the original formulation, the Killing  vector field $\bar{\xi}^\mu$ was merged into the definition of the $(n-2)$-forms $\mathcal{F}_{\mu\nu}$.  
The general formula for the Abbott-Deser charge in Eq. (\ref{general-formula1}) is valid for a broad range of gravitational models, and  an explicit form of the tensor $\mathcal{F}_{\mu\nu}$ is determined by the gravitational model at hand. The present discussion  is confined to the Einstein vacuum equations with a cosmological constant and it is explicitly shown that it can be constructed by using a Thirring superpotential in connection with a background Killing vector field.

On the other hand, in our  approach, $\bar{\xi}_a$ are not merged into the definition of the 2-form to start with, but an expression of the form
$\bar{D}\bar{*}\bar{F}^{a}$ where $\bar{F}^{a}=\frac{1}{2}\bar{F}^{a}_{\fant{a}bc} \bar{e}^{bc}$ are  vector-valued $(n-2)$-forms is obtained first. Subsequently,  the conserved charges are constructed by multiplying the expression by a background Killing vector field to convert the covariant exterior derivative into an exterior derivative. The explicit expression for the 2-forms $\bar{F}^{a}$ can be obtained only after an explicit expression for $*E^a$ is obtained. In the construction of the currents $Q(\xi)$, the crucial role  played by a Killing vector field is to render the energy-momentum complex a scalar quantity
so that the Stokes' theorem can be invoked. In other words, because the forms $\bar{F}^a$ are not  tensorial quantities,  pseudo-tensorial
quantities like $*F^a$ that appear in the total energy expression will not be well  defined, if more than one chart is required for the definition.

Moreover, in a practical computation of Killing energy, a timelike vector field of the form $\xi^\mu=(-1, 0, 0, \cdots)$ is used, which essentially enters the energy expression  to single out the timelike component of the 2-forms $\mathcal{F}^a$. In this regard, the construction of the Killing energy for Einstein field equations, in particular the 2-forms $\mathcal{F}^a$, is related to Thirring's construction of the 2-forms $F^a$ for asymptotically flat spacetimes arising from his attempts to rewrite  the Einstein field equations in a form akin to Maxwell's equations, explicitly in the form as close to the expression $d*F=*J$ as possible \cite{wallner-thirring2,dereli-tucker-QGR}. These considerations lead to the fact that the second step in the construction of the Killing energy is tantamount to rewriting the gravitational field equations as $*E^a=0$ where
\be\label{general-split}
*E^a
=
d*{F}^a-*t^a,
\ee
thus singling out  exact $(n-1)$-forms with all the additional  nonlinear terms pushed  into $*t^a$.
Note that, ${F}^a$ contains the term linear in the derivatives of the metric tensor without resorting to an approximation, 
determining  the metric tensor  perturbatively in a given curved background.

\section{The construction of the Killing Energy}

The spin connections and differential forms are used in \cite{cebeci-sarioglu-tekin1}  to formulate the AD charges related to vacuum Einstein field equations
with a cosmological constant. Namely, for the theory governed by the field equations
\be\label{einstein+cc}
*E^a=-\frac{1}{2}\Omega_{bc}\wdg *e^{abc}+\lambda *e^a=0.
\ee
Using the linearized expressions we derived above for the curvature 2-forms, the Einstein metric field equations (\ref{einstein+cc}) linearize to
\begin{eqnarray}
&&(*G^c+\lambda*e^c)^L
=
-
\frac{1}{2}\left(
\bar{\Omega}_{ab}-\frac{\lambda\bar{e}_{ab}}{(n-2)(n-1)}
\right)\wdg
\bar{*}\bar{e}^{abc}
\nonumber\\
&&
-
\frac{1}{2}\left(
\bar{\Omega}_{ab}-\frac{\lambda\bar{e}_{ab}}{(n-3)(n-2)}
\right)
\wdg
\phi_d\wdg
\bar{*}\bar{e}^{abcd}
\nonumber\\
&&+
\frac{1}{2}
\bar{D}
\left[\left(
\bar{i}_a \bar{D}\phi_b\right)
-
\left(\bar{i}_b \bar{D}\phi_a\right)
\right]
\wdg \bar{*}\bar{e}^{abc}.\label{einstein-eqn-expansion}
\end{eqnarray}
In the linearized expression in Eq. (\ref{einstein-eqn-expansion}), the terms involving the constant parameter $\lambda$ show up both in the zeroth order and in the first order terms. Evidently, it is not possible to have the zeroth order $\lambda$ terms and the first order $\lambda$ terms in the second line on the right hand side of Eq. (\ref{einstein-eqn-expansion}) vanish at once by choosing a particular value of $\lambda$. To proceed further, one assumes that the background Einstein field equations for vacuum with a cosmological constant are satisfied identically. Therefore, by assumption, the parameter $\lambda$ is  fixed by the background curvature and  is not a parameter to have an expansion as in Eq. (\ref{coframe-expansion-def}) for the basis coframe 1-forms \cite{cebeci-sarioglu-tekin1}.  By setting the  zeroth order terms in Eq. (\ref{einstein-eqn-expansion}) equal to zero, one determines a particular value for $\lambda$, and  subsequently use it to replace the background curvature expressions in the first order terms.  Thus, one ends up with the following first order expression:  
\be
(*G^a+\lambda*e^a)^L
=
\frac{2\lambda }{(n-1)}\phi_b
\wdg
\bar{*}\bar{e}^{ab}
+
\bar{D}\left(\bar{i}_b \bar{D}\phi_c\wdg \bar{*}\bar{e}^{abc}\right).
\ee
Eventually, the only unsuitable term that hinders the definition of a flux integral is the first term that appears on the right hand side.

In order to  obtain a flux integral from the expression on the right-hand side, one saturates the free index of the equation with a covariant index of a 
background Killing vector field $\bar{\xi}_c$. Furthermore, if the background Killing vector field is also Killing vector field to all orders, then  the  expression on the left-hand side can be rewritten in the form
\be
\bar{\xi}_a(*G^a+\lambda*e^a)^L
=
\left(\xi_a\left(*G^a+\lambda*e^a\right)\right)^L,
\ee 
using the fact that $\xi^a$ is a Killing vector field.

Commuting  $\bar{\xi}_a$ with the background exterior covariant derivative 
by using the identity below satisfied by the covariant exterior derivative
\begin{eqnarray}
\bar{\xi}_{a}\bar{D}\left[ \bar{i_b}\bar{D}\phi_c\wdg \bar{*}\bar{e}^{abc}\right]
\!\!\!\!&&=
-
\bar{D}\bar{\xi}_{a}\wdg \bar{i_b}\bar{D}\phi_c\wdg \bar{*}\bar{e}^{abc}
\nonumber\\
&&+
\bar{D}\left[
\bar{\xi}_a \left(\bar{i}_b\bar{D}\phi_c\right)\wdg \bar{*}\bar{e}^{abc}
\right],
\end{eqnarray}
one obtains
\begin{eqnarray}
&&\bar{\xi}_a(*G^a+\lambda*e^a)^L
=
d\left(\bar{\xi}_a\bar{e}^b *\left(\bar{D}\phi_b\wdg \bar{*}\bar{e}^{a}\right)\right)
\nonumber\\
&&+
\frac{2\lambda }{(n-1)}\bar{\xi}_a\phi_b
\wdg
\bar{*}\bar{e}^{ab}
-
\bar{D}\bar{\xi}_{a}\wdg \bar{i_b}\bar{D}\phi_c\wdg \bar{*}\bar{e}^{abc}. \label{recuced-eqns-28}
\end{eqnarray}
It is possible to replace now the background covariant exterior derivative on the right-hand side
by an ordinary exterior derivative since the expression  has no free index left.
Furthermore, by making use of the identity
\be
\bar{D}^2\bar{\xi}^a
=
\frac{2\lambda \bar{\xi}_b \bar{e}^{a}\wdg \bar{e}^{b}}{(n-1)(n-2)},
\ee
plus the Killing equation (\ref{killing-eqn}) written in the form 
\be\label{id-c}
\bar{i}_a\bar{D}\bar{\xi}_b+\bar{i}_b\bar{D}\bar{\xi}_a=0,
\ee
and the following identity that can be derived from (\ref{id-c}),
\be
\bar{D}\bar{\xi}_a
=
-\frac{1}{2}\bar{i}_ad\tilde{\bar{\xi}},
\ee
the linearized expression in (\ref{recuced-eqns-28}) takes the remarkably simple  form
\be\label{mrcl}
\bar{\xi}_a(*G^a+\lambda*e^a)^L
=
d\left[
\bar{e}^a \wdg\bar{*}
\left(
\bar{D}\phi_a\wdg \bar{\tilde\xi}
-
\frac{1}{2}
\phi_a\wdg d\tilde{\bar{\xi}}
\right)
\right].
\ee
Here $\tilde{\bar\xi}$ stands for the background Killing 1-form associated with the background Killing vector field $\bar{\xi}=\bar{\xi}^\mu\bar{\partial}_\mu$.

The closed form term on the right-hand side of Eq. (\ref{mrcl}), is a compact way of writing  the expression 
given in Eq. (11) derived in \cite{cebeci-sarioglu-tekin1} that is  mathematically  equivalent to (\ref{mrcl}).

The significance of the AD charges $Q^a(\bar{\xi})$ defined by the flux integral 
\begin{eqnarray}
Q(\bar{\xi})
&=&
\frac{1}{4\Omega_{n-2}G_n}
\int_{U}
d\left[
\bar{e}^a \wdg\bar{*}
\left(
\bar{D}\phi_a\wdg \tilde{\bar{\xi}}
-
\frac{1}{2}
\phi_a\wdg d\tilde{\bar{\xi}}
\right)
\right]
\nonumber\\
&=&
\frac{1}{4\Omega_{n-2}G_n}
\int_{\partial U}
\bar{e}^a \wdg\bar{*}
\left(
\bar{D}\phi_a\wdg \tilde{\bar{\xi}}
-
\frac{1}{2}
\phi_a\wdg d\tilde{\bar{\xi}}
\right)\label{AD-charge-def}
\end{eqnarray}
using the Stokes' theorem  is now apparent from the integrand  in expression $Q(\bar{\xi})$  in Eq. (\ref{mrcl})  that one recovers 
the  expression for the flat background (cf. Eq. (\ref{lin-thirring-form-flat-bck}) below). Note that it is not possible to make the identification with the corresponding expression presented in Eq. (11) in \cite{cebeci-sarioglu-tekin1} for the total gravitational energy.

The well known expression for ADM energy for an asymptotically flat background \cite{ADM} can be obtained from the above expression as well.
In the light of these connections, we find it appropriate to name the expression in Eq. (\ref{mrcl}) the  generalized Thirring form.

We also note that as in the formulation above, the use of Stokes' theorem requires the use of a $p-$form and it is not applicable for the tensor valued forms
that are not tensorial objects in a mathematically strict sense. In addition, the integrand in the flux integral is the canonical injection of the 
expression on the boundary $\partial U$ in a  mathematically strict sense with the orientation adopted from $U$.

The total energy definition in Eq. (\ref{AD-charge-def}) is the main result of the present paper. In computations of the total energy  for illustrative cases, for instance, for the AdS solitons or the Eguchi-Hanson instantons, or for the Taub-NUT-Reissner-Nordstr\"om  solution \cite{cebeci-sarioglu-tekin1}; the term that involves $d\tilde{\bar{\xi}}$ vanishes identically since in these cases, the timelike Killing vector field adapted to the computation has  components $\bar{\xi}^\mu=(-1,0,0\cdots)$. Thus, the calculation of the energy expression derived from the AD recipe partly simplifies  to the calculation of the first term  in the general expression (\ref{AD-charge-def}).

\section{Concluding comments}

The main result of the paper establishes a connection between two different energy definitions both of which use  flux integrals
to define the total energy. In this regard, we may view the expression (\ref{AD-charge-def}) as the generalization of the Thirring 2-forms
to a curved background.

For $\lambda=0$,  the Einstein form linearized around a flat background can be expressed in terms of the Thirring 2-forms
by considering the expression \cite{wallner-thirring2}
\be\label{split}
*G^a=d*F^a-*t^a
\ee
where $*t^a$ in a coordinate coframe corresponds to the Landau-Lifschitz pseudo-energy-momentum tensor and
and the exact expressions for the terms on the right hand side of Eq. (\ref{split}) are given by
\be\label{split2}
*F^a
=
-
\frac{1}{2}\omega_{bc}\wdg *e^{abc}
\ee
and
\be
*t^a
=
\frac{1}{2}
(\omega_{bc}\wedge\omega^{a}_{\phantom{Q}d}
\wedge*e^{bcd}
-
\omega_{bd}\wedge\omega^{d}_{\fant{a}c}\wedge
*e^{abc}).
\ee
As a consequence of the splitting (\ref{split}), one explicitly  
has 
\be\label{lin-thirring-form-flat-bck}
(*G^a)^L
=
(d*F^a)^L
\ee
where orthonormal coframe basis $\{\bar{e}^a\}$ stands for the natural Cartesian basis $\{dx^a\}$ for the flat background.
Using the perturbation 1-forms $\phi_a$ and assuming the flat background in the linearization formula discussed in the previous sections,
the expression for the Thirring 2-form (\ref{split2}) can be reduced to \cite{baykal-dereli-epj-plus}
\be\label{split3}
\bar{*}\bar{F}^a
=
\bar{e}^b\wdg \bar{*}(d\phi_b\wdg \bar{e}^a)
\ee
where now the barred quantities refer to flat background. The generalization  of the expression (\ref{split3}) to a curved background 
is simply obtained by replacing the exterior derivatives with background covariant exterior derivatives.
The linearized expressions (\ref{split2}) and (\ref{split3}) are to be compared with curved background expressions  
linearized (\ref{gen-thirring-form1}) and (\ref{gen-thirring-form2}), respectively.

The total energy for an asymptotically flat spacetime then follows from the integral of the timelike component 
of the Thirring 2-forms defined even in the form given in Eq. (\ref{split2}) over $S^{(n-2)}$ in the limit the radius goes to infinity that filters out the asymptotic behavior of the coframe metric components arising from the infinite radius limit. 

For the generalized expression (\ref{AD-charge-def}), the curved background assumption brings out additional complications  regarding the boundary
of the flux integral. For instance, because of the presence of the horizons for a dS background, the boundary $\partial U$ is to be introduced carefully avoiding a horizon in order to define a flux integral to have a well defined value, whereas for the AdS background  there is no such restriction \cite{abbott-deser}.

The construction of the Abbott-Deser charges for the particular case of the Einstein vacuum equations with a cosmological constant using a Thirring superpotential  offers further insight, and facilitates the comparison of energy definitions for more involved gravitational models as well. 
In addition, the above construction  has the technical advantage that it simplifies the explicit tensorial  components of the closed form $\mathcal{F}_{\mu\nu}$ involved in some other cases of interest which can usually be obtained after some tedious calculations (See the discussion of this issue in Ref. \cite{cebeci-sarioglu-tekin1} following Eq. (7) therein). For example, tor the higher curvature gravitational models which admit  maximally symmetric solutions, the expression of the form in Eq. (\ref{split}) can readily be used. More precisely, the determination of an exact $(n-1)$-form from the metric equations of such a higher curvature model can be carried out after the field equations are derived using the first order formalism with the vanishing torsion and nonmetricity contraints  \cite{dereli-tucker-var}-\cite{hehl} from a coframe variational derivative. As a particular  example, one can show that the linear terms arise from the Lagrange multiplier term  which imposes the zero torsion condition for the general quadratic curvature model \cite{baykal-qc-energy}. For the general quadratic curvature model, the related exact form  is bound to  be an expression involving the exterior derivatives and the background Hodge duals of the 2-forms $F^a$ and its contraction for the flat background. The corresponding expression for the curved background is simply obtained by replacing exterior derivatives with the background covariant exterior derivative. In this regard, the Einstein field equations (\ref{einstein+cc}) can be considered as an exceptional case since the Lagrange multiplier vanishes identically for the Einstein gravity with a cosmological constant.

The proof of the positivity of  energy in Einstein's theory of gravity by Witten \cite{witten} that uses spinors 
necessitates the introduction of an orthonormal coframe and the spin connection formalism, and in particular the Thirring 2-forms \cite{straumann}. In this context, the identification of Abbott-Deser energy in terms of  Thirring 2-forms can be technically helpful in similar considerations for higher curvature models.

\acknowledgments

We would like to thank the anonymous Referee for the constructive comments and for pointing out the typos.

\end{document}